\documentclass[runningheads]{llncs}
\usepackage{esvect}
\usepackage[T1]{fontenc}
\usepackage{cite}
\usepackage{amsmath,amssymb,amsfonts}
\usepackage{algorithmic}
\usepackage{graphicx}
\usepackage{textcomp}
\usepackage{xcolor}
\usepackage{uppaal}
\usepackage{subcaption}
\def\BibTeX{{\rm B\kern-.05em{\sc i\kern-.025em b}\kern-.08em
    T\kern-.1667em\lower.7ex\hbox{E}\kern-.125emX}}

\newcommand{\DRODS}{\textit{D-RODS}}
\def\NN{\mathcal{N}}
\def\R{\mathbb{R}}
\def\haty{\hat{y}}
\def\hatx{\hat{x}}
\def\hatB{\hat{B_T}}
\def\hatBo{\hat{Bo_T}}
\def\hatWM{\hat{W_M}}
\def\hatWoM{\hat{Wo_M}}
\def\preceqcomp{\preceq}

\begin{document}

\title{Contract-based Verification of Digital Twins}

\author{Muhammad Naeem \and
Cristina Seceleanu}
\authorrunning{F. Author et al.}
\institute{School of Innovation, Design, and Engineering, \\ Mälardalen University, Västerås, Sweden\\
\email{\{muhammad.naeem, cristina.seceleanu\}@mdu.se}}

\maketitle

\begin{abstract}
Digital twins are becoming powerful tools in industrial applications, offering virtual representations of cyber-physical systems. However, verification of these models remains a significant challenge due to the potentially large datasets used by the digital twin. 
This paper introduces an innovative methodology for verifying neural network-based digital twin models, in a black-box fashion, by integrating model checking into the process. The latter relies on defining and applying system-level contracts that capture the system's requirements, to verify the behavior of digital twin models, implemented in Simulink.
We develop an automated solution that simulates the digital twin model for certain inputs, and feeds the predicted outputs together with the inputs to the contract model described as a network of timed automata in the \uppaal model checker. The latter verifies whether the predicted outputs fulfill the specified contracts. This approach allows us to identify scenarios where the digital twin's behavior fails to meet the contracts, without requiring the digital twin's design technicalities. 
We apply our method to a boiler system case study for which we identify prediction errors via contract verification. Our work demonstrates the effectiveness of integrating model checking with digital twin models for continuous improvement. 
\end{abstract}
%%%%%%%%%%%%%%%%%%%%%%%%%%%%%%%%%%%%%%%%%%
\section{Introduction}
\label{intro} The emergence of Industry 4.0 has propelled Digital Twins (DT) to the forefront of industrial digitalization, offering virtual representations of cyber-physical systems that facilitate precise simulations, analysis, and control \cite{tao2018digital}. As industrial systems grow in complexity, ensuring the reliability and correctness of these DT becomes crucial. 

This work is a part of the Dynamic and Robust Distributed Systems (D-RODS) project aiming to develop an adaptable and dependable framework for the implementation and operation of DT, tackling issues in industrial digitalization including system integration, performance enhancement, and compatibility with legacy systems \cite{seceleanu2023building,gu2024service}.
In the current study, we focus on the Verification and Validation (V\&V) of Neural Network-based Digital Twins (NNDT) models using model-checking techniques. NNDT have the potential to predict system behavior, conduct real-time tests, and identify optimization opportunities. However, their effectiveness depends on their ability to represent and accurately respond to real-world conditions \cite{grieves2017digital}.

Recent advancements in model-driven engineering, such as the CoCoSim framework, have demonstrated the value of integrating formal verification techniques with model-based design, especially for multi-periodic discrete systems \cite{bourbouh2020cocosim}. %Similarly, the application of design by contract principles to model execution verification has shown promise in ensuring the correctness of model behavior during runtime \cite{cariou2011contracts}.

Our research applies a similar concept to the domain of DT, proposing a novel approach that employs model checking to verify monotonicity, functional, and infrastructure contracts specified for black-box NNDT. We propose a methodology for systematically verifying the behavior of DT models implemented in \texttt{Simulink} by defining system contracts, as networks of \uppaal timed automata~\cite{behrmann2006tutorial}.
We present an automated solution that connects a simulation environment with model checking, allowing for the continuous validation and improvement of NNDT models. Our approach detects situations where the NNDT generates incorrect outcomes.

Due to their dynamic nature and complex interactions with cyber-physical systems, verifying DT models is challenging. Dahmen et al. emphasize the importance of systematic methods to verify and validate DT models \cite{dahmen2022verification}, and propose an approach that breaks down the verification problem into smaller, more manageable components. This research aligns with our methodology of using (sub-)system contracts to verify specific aspects of DT behavior.

Given the importance of ensuring the correctness of DT behavior, our research aims to address the following research question: \textbf{RQ:} How can we design an automated solution to model check a black-box NNDT model?

To answer it, this paper presents a contract-based approach for verifying black-box NNDT models in \uppaal. The solution simulates the \texttt{Simulink} NNDT model for certain inputs, and feeds the predicted outputs together with the inputs to the contract model described as a network of timed automata in the \uppaal model checker. The latter verifies if the predicted outputs fulfill the specified contracts. In brief, our contributions are as follows:
\begin{itemize}
    \item \textit{Model-Checking-based Methodology:} We propose a methodology that uses contracts modeled as \uppaal timed automata, and employs model checking to verify the correctness of black-box NNDT. 
    \item \textit{System Contract Implementation:} We define a systematic approach for modeling system contracts that specify expected behaviors, enabling precise verification of model outputs against system requirements. 
 %
 %\item \textcolor{gray}{\textit{Automated Feedback Loop:} We have implemented an automated feedback mechanism that integrates verification results into the refinement process. This allows for iterative improvements in the accuracy and reliability of DT models.}
%
    \item \textit{Practical Validation:} Our approach is illustrated through a burner-boiler system case study, showcasing its effectiveness on an industrial application.
\end{itemize}

%This research contributes to the evolving landscape of Industry 4.0 by addressing the critical need for robust verification methods in DT development. 
Our approach facilitates the identification of scenarios where the DT's behavior fails to meet the contracts, without requiring knowledge of the model's internal structural and functional details. 
%We develop an automated solution that simulates the DT model for specific inputs and uses the \uppaal model checker to verify if the predicted outputs fulfill specified contracts. This method enables us to identify scenarios where the DT's behavior fails to meet the contracts and suggest mitigation solutions in the learning process.

The remainder of this paper is organized as follows.
Section \ref{Preliminaries} provides the necessary background information. Section \ref{Mthd} details our methodology for verifying NNDT models using contract-based model checking. Section \ref{UC} describes our case study of a burner-boiler system, including the DT model implemented in \texttt{Simulink}, as well as the system's contracts. Section \ref{Modelling} presents the implementation of contracts in \uppaal, for model checking, and Section \ref{verification} discusses the verification results. Finally, Section \ref{RW} presents and compares to relevant related work, before Section \ref{conclusion} concludes the paper, summarizing our contributions and discussing potential lines of future work.
%%%%%%%%%%%%%%%%%%%%%%%%%%%%%%%%%%%%%
\section{Preliminaries}
\label{Preliminaries}
This section presents the background information necessary to understand the concepts and methodologies employed in our study. 
\subsection{The \DRODS\ Approach} \label{sec.DRODS_Approach}
The \DRODS\ project, proposed a cutting-edge approach for development and operation of DT. This innovative framework aims to address the complexities inherent in industrial digitalization by integrating artificial intelligence (AI) and formal verification techniques. 

\noindent \textbf{Context: Physical (CP)} represents the physical components including machineries and controllers. 

\begin{figure*}%[htbp]
    \begin{center}
    \includegraphics[width =\linewidth]{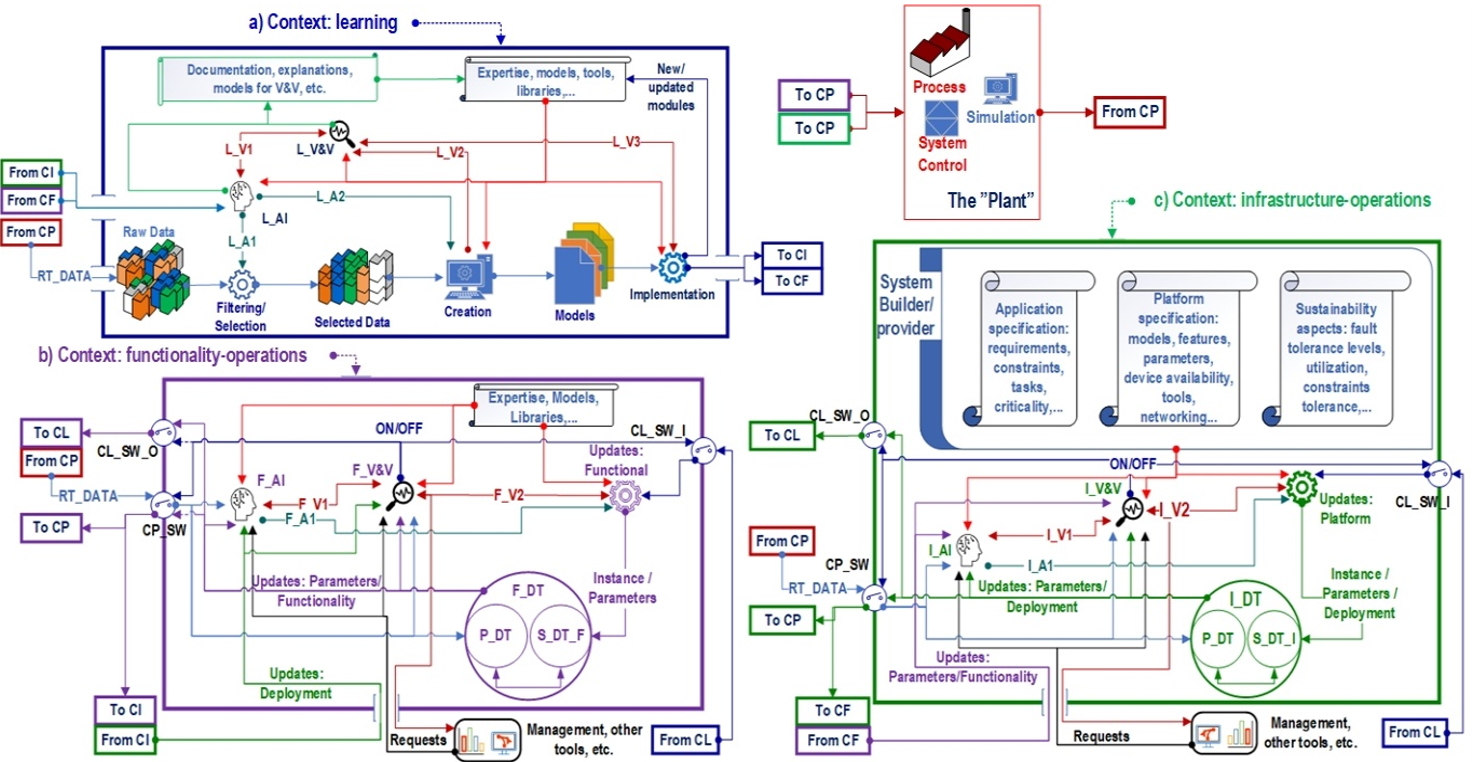}
    \caption {D-RODS approach details}
    \label{fig.D-RODS_Approach_details} %RENAME \label{fig:nodeTemplate}
    \end{center}
    \end{figure*}

\noindent \textbf{Context: Learning (CL)} layer (Fig.\ref{fig.D-RODS_Approach_details}a)) focuses on creating DT models through unsupervised learning. These models are derived from extensive data histories and existing domain expertise.
    
\noindent \textbf{Context: Functionality (CF)} and \textbf{Context: Infrastructure (CI)} are similar layers (Fig.\ref{fig.D-RODS_Approach_details}b, c), each containing AI and V\&V components. These contexts supervise and enhance the development and execution of complex DT, facilitating continuous learning, optimization, and behavior evaluation.

At its core, \DRODS\ seeks to unite DT, AI, and V\&V technologies in a novel architectural setup. This integration aims to increase the trustworthiness of AI approaches through formal verification and analysis while optimizing operations, resource utilization, and power consumption. The framework is designed to support high levels of autonomy by ensuring the accuracy and efficiency of employed models through continuous learning and verification.

\DRODS\ is committed to improving system performance forecasting and optimizing resource allocation in the field of AI. The project also aims to integrate formal verification and runtime testing to ensure the accuracy of developed models and continuously monitor operational correctness. Through these advancements, \DRODS\ aims to advance DT technology, offering a comprehensive framework that addresses the challenges of complex industrial systems while promoting efficiency, reliability, and adaptability.

\subsection{\uppaal Timed Automata}
\label{MC}
Model checking is a formal verification technique that systematically and exhaustively verifies whether a system model meets specified requirements. \uppaal~\cite{behrmann2006tutorial} is a state-of-the-art model checker designed for modeling, verification, and simulation of distributed systems. Its modeling framework is based on Timed Automata (TA), which extends traditional finite-state machines by incorporating real-valued clocks. These clocks allow for the precise timing of events, enabling the modeling of systems where timing is critical.

In \uppaal, the concept of TA is further enhanced through the use of \uppaal Timed Automata (UTA). UTA introduces discrete data variables, such as integers and Boolean values, which can act as guards or expressions to control transitions between states. Additionally, UTA supports specifications in the C programming language, facilitating the observation and control of the model's discrete states.

A UTA can be formally defined as a tuple: $\langle L, l_{0}, C_h, V, E, I \rangle$, where: $L$ is a finite set of locations in the automata model, $l_{o}$ represents the initial location,
$C_{h} = C_{h}!\, \cup \,C_{h}?$ denotes a set of channels for communication, with $C_{h}!$ and $C_{h}?$ representing sending and receiving channels, respectively~\cite{naeem2024energy}.
$V$ includes a set of data variables and clocks,
$E$ consists of edges connecting locations,
$I$ specifies invariants that must hold true for certain expressions.

\begin{figure}%[htbp]
  \centering
  \begin{subfigure}{.45\linewidth}
    \centering
    \includegraphics[width = \linewidth]{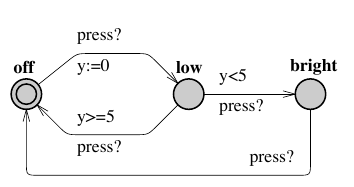}
    \caption{Lamp}
  \end{subfigure}%
  \hspace{+0.5em}
  \begin{subfigure}{.23\linewidth}
    \centering
    \includegraphics[width = \linewidth]{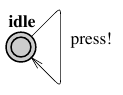}
    \caption{User}
  \end{subfigure}%
  \caption{The simple lamp example.}
  \label{UPPAAL_Examples}
\end{figure}

\uppaal 's ability to model and verify real-time systems makes it particularly valuable in safety-critical domains. The tool offers a user-friendly graphical interface for designing models and simulating their behavior, along with a command-line interface for efficient verification processes~\cite{naeem2021battery}.

%Recent developments in \uppaal have expanded its functionality to include statistical model checking and support for probabilistic timed automata. These enhancements allow for more comprehensive analyses of complex systems that involve uncertainty. 

Fig \ref{UPPAAL_Examples}(a) shows a {\em network of UTA} modeling a simple lamp and its user~\cite{naeem2023modelling}. 
The lamp has three locations: off, low, and bright. If the user presses a button, i.e., synchronizes with the press? then the lamp is turned on. If the user presses the button again, the lamp is turned off. However, if the user is fast and rapidly presses the button twice, the lamp is turned on and becomes bright. The user model is shown in Fig \ref{UPPAAL_Examples}(b). The user can press the button randomly at any time or even not press the button at all. The clock y of the lamp is used to detect if the user is fast $(y < 5)$ or slow $(y >= 5)$.

The contracts (queries in UPPAAL) to be verified by model checking on the resulting network are specified in a decidable subset of (Timed) Computation Tree Logic ((T)CTL). In this paper, we verify (T)CTL queries of the following kinds ($p$ is a state property): (i) {\sf Reachability:} $E\, \lozenge\, p$ - the query evaluates to true if there exists a path where $p$ eventually holds, and (ii) {\sf Invariance:} $A\, \Box \, p$ - the query evaluates to true if (and only if) every reachable state satisfies p, in other words, for all paths $p$ always holds.  
%%%%%%%%%%%%%%%%%%%%%%%%%%%%%%%%%%%%%
\section{Proposed Methodology}
\label{Mthd}
In this section, we describe our methodology for verifying NNDT models in a black-box fashion, by integrating model checking into the process. %Our approach allows for verification without requiring knowledge of the DT's internal functional details, making it particularly valuable for complex industrial applications.
\vspace{-.2cm}
\subsection{Digital Twin Model as a Neural Network} 
The starting point of our methodology is a black box NNDT model, implemented in \texttt{Simulink} (in this paper). However, the approach can be applied to other modeling frameworks, straightforwardly. The NNDT serves as a virtual representation of the cyber-physical system, continuously updated with real-time data to mirror its current state and behavior.

%Our DT model takes various inputs such as fuel flow rate, air intake, and current temperature readings. It then processes this data through its neural network architecture to output predictions of future system states, including boiler temperature, pressure, and efficiency metrics.

The black-box nature of this neural network-based model allows us to capture complex, non-linear relationships within the system without requiring detailed knowledge of its internal structure. This approach provides flexibility and accuracy in representing the system's behavior across various operating conditions. The inputs and outputs form the basis for our subsequent contract verification process, allowing us to assess the NNDT's performance against predefined system contracts.
\subsection{Contract Model Development} \label{cmd}
We develop a comprehensive contract model using UTA, and provide contract verification in \uppaal. These contracts capture the system's requirements and serve as a formal specification of the expected behavior of the NNDT. We categorize the contracts into three main types:
\begin{figure*}[t]
    \begin{center}
    \includegraphics[width =\linewidth]{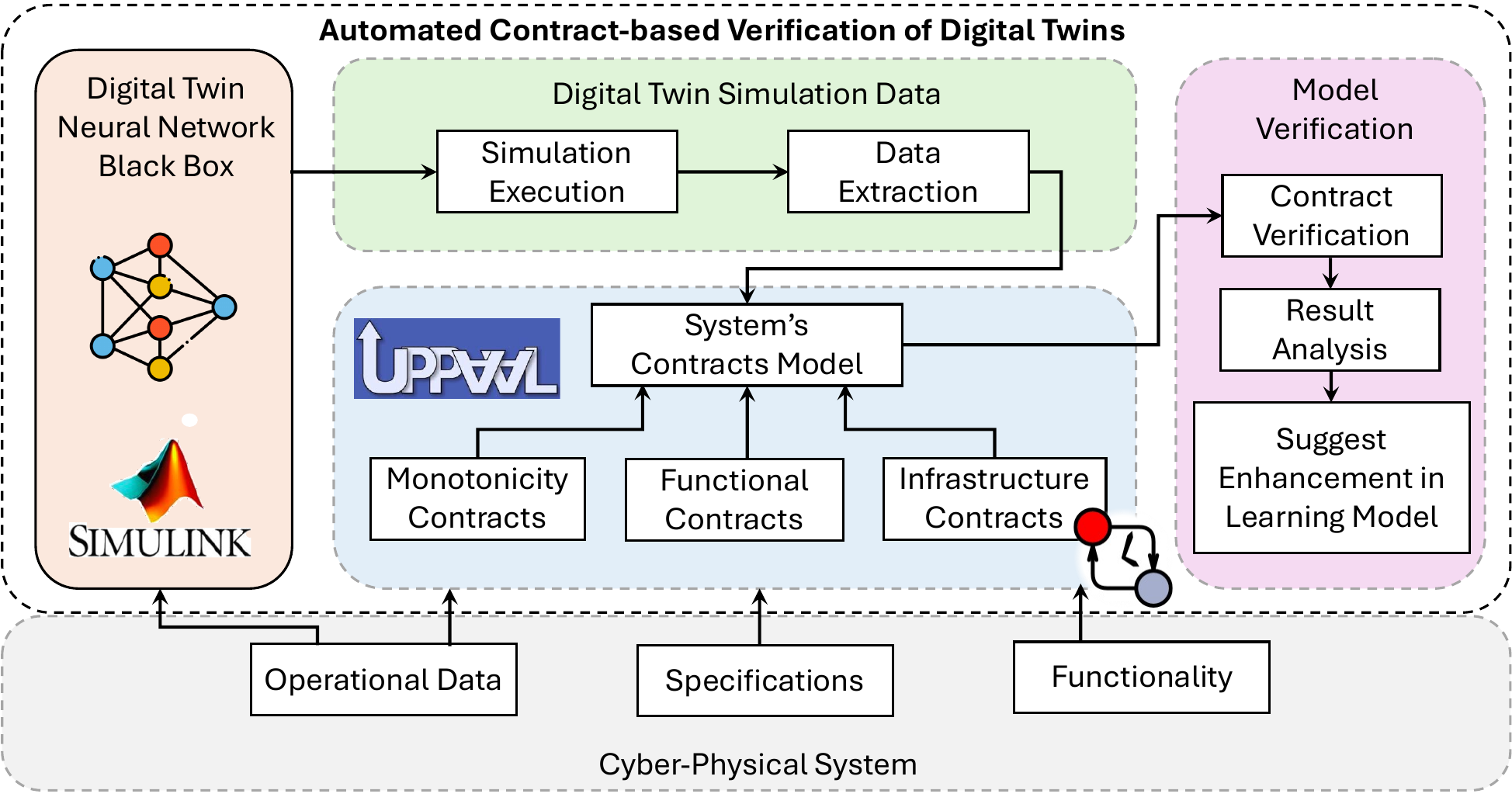}
    \caption {Methodology}
    \label{fig.Methodology}
    \end{center}
    \end{figure*}
%\begin{description}

\begin{itemize}
    \item \textit{Monotonicity Contracts:} These contracts are designed to verify that the output of the NNDT behaves monotonically with respect to certain inputs. This means that if an input's value increases, the NNDT's output value should also increase, and vice versa. This property is crucial in many industrial applications where a predictable relationship between inputs and outputs is expected. 
    \item \textit{Functional Contracts:} These contracts specify the expected functional behavior of the digital twin, defining relationships between inputs and outputs that must be maintained throughout the operation of the model. These contracts ensure that the digital twin follows the intended operational specifications.
    \item \textit{Infrastructure Contracts:} These contracts focus on the operational aspects of the digital twin, such as response times, resource usage, and system stability. These contracts ensure that the digital twin operates within acceptable performance limits.
\end{itemize} 

\paragraph{{\bf{Contracts for Neural Networks-based Digital Twins (NNDT).}}} In the following, we define our notion of {\em NNDT contract}, as used in this paper. 
\begin{definition}[NNDT Contract]\label{NNcontract}
Let $\NN: \R^n \to \R^m$ be a digital twin represented by a neural network, mapping an input vector $x \in \R^n$ to an output vector $y \in \R^m$. We define the {\em assume-guarantee} contract for the digital twin, as the pair (A, G), as follows: 
\begin{itemize}
    \item {\rm{Assumption (A)}}: We assume that the input and output of the digital twin at time $t$, denoted as $x(t)$, and $y(t)$, respectively, are represented by a moving average over the last $m$ time steps ($m$ is a constant that depends on the application), to mitigate the effects of inaccurate sensing, and predictions, respectively:
\begin{equation}
    \hatx(t) = \frac{1}{m} \sum_{i=1}^{m} x(t - i),
\end{equation}
where $\hatx(t)$ is the stabilized input, and:
\begin{equation}
    \haty(t) = \frac{1}{m} \sum_{i=1}^{m} y(t - i),
\end{equation}
where $\haty(t)$ is the stabilized output. To these cross-cutting assumptions (they hold for all contracts), we add predicate $p$ that captures the requirements on inputs $x(t)$. 
    \item {\rm{Guarantee (G)}}: A predicate $q$ that captures the NNDT's expected output behavior. As an example, a {\em monotonicity contract} that states that, if the input $\hatx$ increases component-wise, then the output of the neural network, $\NN(\hatx)$, should also increase component-wise, can be expressed formally as follows (similarly, for the decreasing case): 
\begin{equation}
    ((\forall \hatx, \hatx' \in \R^n .\, \hatx \preceqcomp \hatx'), (\NN(\hatx) \preceqcomp \NN(\hatx'))
\end{equation}
Here, $\preceqcomp$ denotes component-wise inequality (that is, inequality that is applied individually to each entry, which can be the average value over $m$ previous ones, as explained above). This ensures that the neural network maintains monotonicity and avoids erratic behavior in response to changes in input. 
\end{itemize}
\end{definition}
The semantics of the (A, G) contract of Definition \ref{NNcontract} is given in terms of {\em \uppaal timed automata} (UTA), as follows. 
\begin{definition}[Semantics of NNDT Contracts] \label{NNDTSemContract}
A contract for a neural-network-based digital twin is a pair (A, G), where $A = ||_i UTA_{A_i}$ represents the network of $i\neq 0$ {\bf\em assumption UTA}, and $G = ||_j UTA_{G_j}$ is the network of $j \neq 0$ {\bf\em guarantee UTA}. The semantic contract is then defined by the parallel composition $A||G$. 
\end{definition}

\subsection{Role of Contracts in Verification}
The contracts mentioned above play a critical role in the verification process because they set clear expectations for how the digital twin should behave. During the model-checking phase, we compare the outputs predicted by the NNDT model to these contracts. If the outputs do not meet any of the contracts, it shows that there might be a problem with how the model is working.

By using these contracts in a structured way, we can ensure that the digital twin meets its required functions and follows the necessary rules for operation. This helps to improve the reliability and safety of the system that it represents.
\vspace{-.2cm}

\paragraph{{\bf{Verifying Satisfaction of NNDT Contracts.}}} In our approach, the verification of NNDT contracts reduces to model checking the UTA parallel composition $A||G$ of Definition~\ref{NNDTSemContract}, against {\em invariance} properties of the form $A\Box\, q$, where $q$ is the predicate that models the guarantee, per component. As sometimes it proves  faster, we can also check {\em reachability} properties of the form $E <> \neg q$, where a witness trace returned by \uppaal identifies a breach of contract. 

\subsection{Automated Verification Process}
We develop an automated solution to bridge the gap between the \texttt{Simulink}-based NNDT and the \uppaal contract model. This solution consists of the following steps:
\begin{itemize}
    \item \textit{Simulation Execution} The automated system triggers simulations of the DT model in \texttt{Simulink} across various input scenarios.
    \item \textit{Data Extraction} The system captures the text output data generated by the \texttt{Simulink} model during these simulations.
    \item \textit{Contract Verification} The captured data is fed into the \uppaal contract model obtained as described above, which is then model checked against the specified contracts, hence verifying the NNDT's behavior compliance.
\end{itemize}
\subsection{Results Analysis } We analyze the verification results, identifying any violations of the contracts, where the NNDT's behavior fails to meet the contracts. This process involves:

\begin{itemize}
    \item Pinpointing the exact input conditions that lead to contract violations.
    \item Categorizing errors based on the type of contract breached (monotonicity, functional, or infrastructure).
\end{itemize}
%%%%%%%%%%%%%%%%%%%%%%%%%%%%%%%%%%%%%
\section{Case Study: Burner-Boiler System}
\label{UC}
\begin{figure}%[htbp]
    \centering
    \includegraphics[width=0.8\columnwidth]{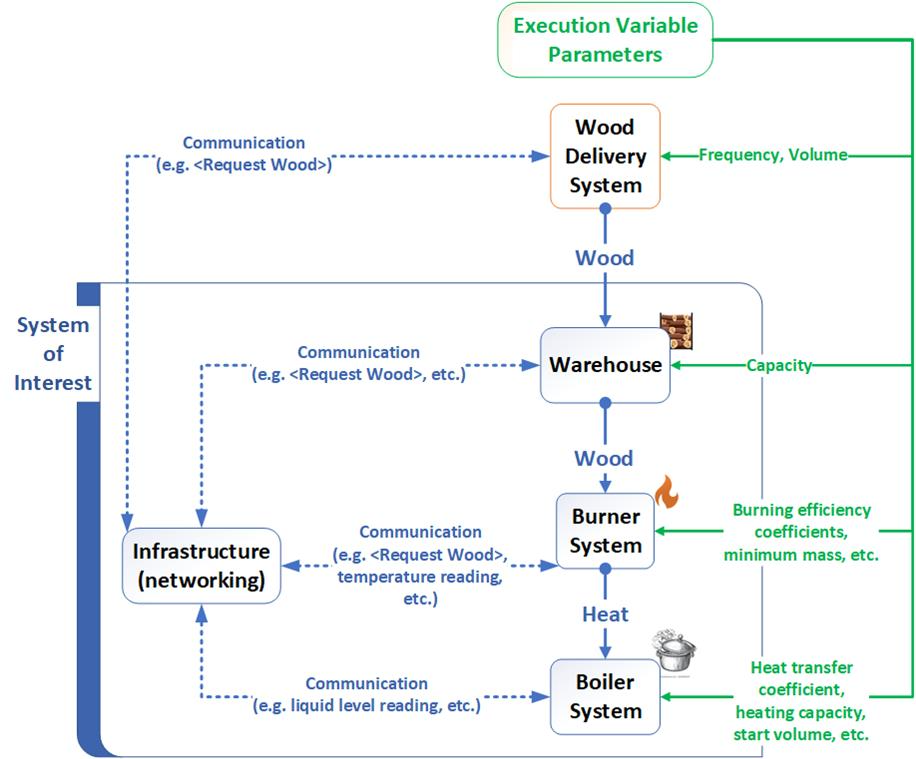}
    \caption{The heating system.}
    \label{fig.UseCase}
\end{figure}

To illustrate our approach, we apply it on a case study focused on heating different liquids until they evaporate (Fig \ref{fig.UseCase}). This model includes several parts that work together to achieve the goal. 

At the center of the system, there is a \emph{wood provider}, which supplies fuel to a \emph{warehouse}. From there, a \emph{burner system} takes the fuel to heat up a \emph{boiler}. The boiler heats containers made of various materials, each holding different types of liquids. The system's complexity comes from several randomly assigned rates. These include how efficiently the wood burns, how well heat transfers in the containers, and how quickly each liquid heats up. Key parameters, such as the starting amounts of wood and liquid, sizes of wood deliveries, and the evaporation temperatures of the liquids, are also randomly set. This randomness allows for generating a variety of scenarios in our study. 

Fig \ref{fig.UseCase} shows a simple diagram of this setup, highlighting the main parts, as well as how information flows through the system. The model is built around the \emph{System of Interest (SoI)}, which includes three main subsystems:

\begin{itemize}
    \item \emph{Warehouse:} The central place for storing and distributing fuel.
    \item \emph{Burner:} Responsible for burning fuel and generating heat.
    \item \emph{Boiler:} The part where liquids are heated.
\end{itemize}

There is also an external \emph{Wood Delivery System} that interacts with the SoI but operates separately. While this system is important for supplying wood, it is not included in our DT modeling. We can change its delivery conditions—like when and how much wood is delivered—but it remains outside our current focus.

Our main goal is to create efficient and accurate DT models. In this paper, we focus on the verification of a black-box NNDT model developed in our previous work. This model represents the complex interactions within the burner-boiler system, and our current efforts aim to ensure its correctness and reliability using \emph{contract-based verification} in the \uppaal tool.
\subsection{Digital Twin Model}
\label{DT}
\begin{figure*}%[htbp]
    \centering
    \includegraphics[width=\linewidth]{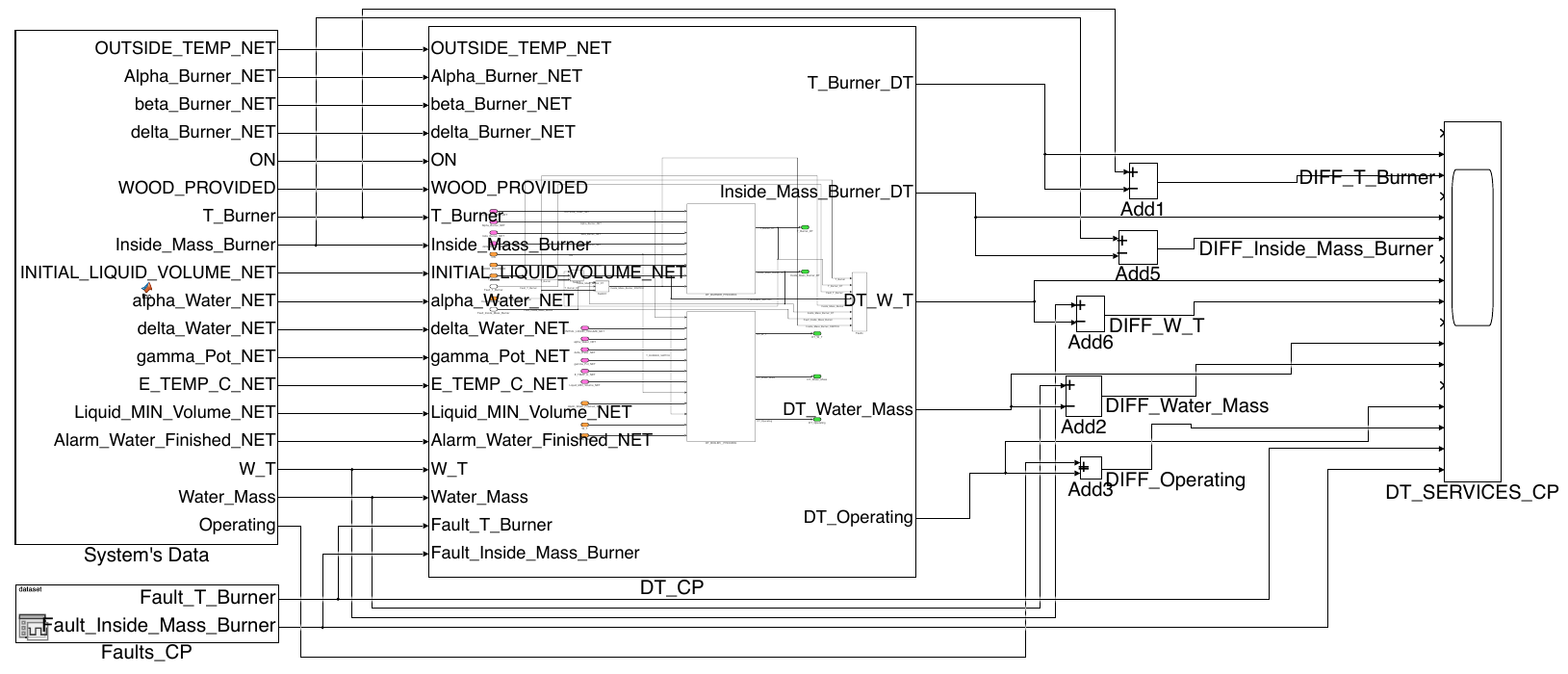}
    \caption{DT Model}
    \label{fig.DigitalTwin}
\end{figure*}

The Fig \ref{fig.DigitalTwin} illustrates the DT model in \texttt{Simulink}.
The NNDT model for the burner-boiler system is implemented in \texttt{Simulink} and serves as a virtual representation of the physical boiler. This model integrates real-time data to accurately reflect the current state and behavior of the boiler.

We use the following notation to describe the DT model for this use case:
\begin{minipage}{0.5\textwidth}
    \textbf{$B_T$:} Burner Temperature  \\
    \textbf{$W_T$:} Water Temperature  \\
    \textbf{$Wo_M$:} Wood Mass  \\
    \textbf{$t$:} Current time step  \\
    \textbf{$W_A$:} Water Alarm   \\
    \textbf{$Wo_A$:} Wood Alarm   \\
    \textbf{$A$:} Assumption   \\
\end{minipage}%
\begin{minipage}{0.5\textwidth}
    \textbf{$Bo_T$:} Boiler Temperature  \\
    \textbf{$W_M$:} Water Mass  \\
    \textbf{$T_{env}$:} Environmental Temperature  \\
    \textbf{$T_{\text{Boil}}$:} Boiling Temperature  \\
    \textbf{$Wo_D$:} Wood Delivered   \\
    \textbf{$Wo_R$:} Wood Request   \\
    \textbf{$G$:} Guarantee   \\
\end{minipage}

The model takes several input parameters, including temperatures, burner characteristics (Alpha, Beta, Delta), wood and water mass and alarms. These inputs allow the DT to adapt to changing conditions and predict the boiler's operations effectively.

Key outputs from the model include predictions for $B_T$, $Bo_T$, $W_M$, $Wo_M$, $W_A$, $Wo_A$, $Wo_R$ and $Wo_D$. The DT uses a neural network to process these inputs and generate accurate predictions about future states of the system.

While the model has been trained on specific datasets and validated with known data, there remains a potential for errors due to limitations in the available data. This aspect is critical as it highlights the scope of our work in this paper. We focus on model-checking the NNDT black-box model using the contract-based approach described in Section~\ref{Mthd}.

Through the use of its neural network architecture, this DT captures complex relationships within the boiler system. Our contract-based verification method enables us to capture specific behavioral requirements for the NNDT, allowing us to assess its contract compliance. This approach ultimately aims to improve the reliability and accuracy of real-world applications NNDT.

\subsection{System Contracts}
We define three categories of contracts for our DT model: \emph{Monotonicity Contracts (MC)}, \emph{Functionality Contracts (FC)}, and \emph{Infrastructure Contracts (IC)}. Each contract is formulated as a pair, $(A, G)$, using the notation described in Section \ref{cmd}. In cases where multiple assumptions exist, we have: $A = \land {_i} A_i$, and similar for multiple guarantees: $G = \land {_i} G_i$.   

\subsubsection{Monotonicity Contracts (MC)}
These contracts are designed to verify that the output of the NNDT behaves monotonically with respect to certain inputs, which is a crucial property of NN. 

\hspace{-0.6cm}\textbf{\textit{MC1 ($B_T$ vs $Bo_T$): }} $A1:\; \forall t, \quad B_T(t) < T_{\text{Boil}}$
\begin{center}
    $A2:\; \hatB(t) > \hatB(t-1)\;\; ,\;\; G1:\; \hatBo(t) \geq \hatBo(t - 1)$

    $A3:\; \hatB(t) < \hatB(t - 1)\;\; ,\;\; G2:\; \hatBo(t) \leq \hatBo(t - 1)$  
\end{center}
     When the burner temperature $\hatB$ is below the boiling point $T_{\text{Boil}}$, it may increase or decrease smoothly over time. The boiler temperature $\hatBo$ must follow the trend of the burner temperature, either increasing or decreasing, accordingly. 

%*******************  MC2
\hspace{-0.6cm}\textbf{\textit{MC2 ($W_M$ vs $Bo_T$): }}
\begin{center}
$A1: \;\hatBo(t) > T_{\text{Boil}}\;\; ,\;\; G1:\;  \hatWM(t) \leq \hatWM(t - 1)$

$A2:\; \hatBo(t) \leq T_{\text{Boil}}\;\; ,\;\; G2:\; \hatWM(t) \approx \hatWM(t - 1)$
\end{center}
When $\hatBo$ exceeds $T_{\text{Boil}}$, evaporation occurs, causing $W_M$ to decrease over time. However, if the $\hatBo$ drops to or below $T_{\text{Boil}}$, the evaporation should slow down or stop, stabilizing the $\hatWM$. This ensures that the system correctly models the evaporation dynamics, maintaining a realistic relationship between $Bo_T$ and $W_M$.

%*******************  MC3
\hspace{-0.6cm}\textbf{\textit{MC3 ($Wo_M$ vs $B_T$): }}$A1:\; \forall t, Burner = ON$ 
\begin{center}
$A2: \;\hatWoM(t) > \hatWoM(t - 1)\;\; ,\;\; G1:\;  \hatB(t) \geq \hatB(t - 1)$

$A3: \;\hatWoM(t) < \hatWoM(t - 1)\;\; ,\;\; G2:\;  \hatB(t) \leq \hatB(t - 1)$
\end{center}
$\hatB$ depends on the $\hatWoM$. When wood burns, its mass decreases. If more wood is added, $\hatB$ should increase or stay the same. If wood mass decreases, $\hatB$ should eventually drop.

By establishing these monotonic relationships, the verification process ensures that the digital twin accurately simulates the boiler system's thermal behavior. Any contract violation suggests issues in the neural network model, data handling, or system design, helping in detecting potential problems.
\subsubsection{Functionality Contracts (FC)}
The functionality contracts define the expected operational behavior of the system, by specifying how different components should respond to certain conditions. 

\textbf{\textit{1- Burner System FC:}}

The {\em burner system} ensures safe and efficient operation by managing wood supply, monitoring temperature, and responding to boiler signals. 

\begin{description}
    \item [\textbf{\textit{FC1: }}] $ A: \;\hatWoM(t) < Wo_{M_{min}}\;\; ,\;\; G:\;  \text{RequestWood}()$
    
    The burner must request wood when the $\hatWoM$ falls below a certain threshold.
\vspace{.2cm}
    \item [\textbf{\textit{FC2: }}] $A1: \;(Wo_R(t) \wedge \neg Wo_D(t+\Delta t) \text{ where } \Delta t = 60\text{s}) \; ,\;A2: (\hatWoM < Wo_{M_{min}})\;\; ,G:\;  W_A$
    
    The $W_A$ should be triggered if no wood is received within 1 minute after the wood request, or when $\hatWoM$ falls below the minimum level.
\vspace{.2cm}
    \item [\textbf{\textit{FC3: }}] $A: \;\hatWoM(t) > Wo_{min}\;\; ,\;\; G1:\;   \neg Wo_R(t) \; ,\;\; G2:\; \neg Wo_D(t)$
    
    This states that there should be no wood request or delivery when the $\hatWoM$ is above $Wo_{min}$.
\vspace{.2cm}
    \item [\textbf{\textit{FC4: }}] $A1: \; ReachedIdealRange(t) = \text{True}\; , \;A2: \;\hatB(t) < 130^\circ C \lor \hatB(t) > 160^\circ C) 
    \;\; ,\;\; G:\;   \text{CriticalTemperatureAlarm}(t)$
    
    The system must enable the \textit{CriticalTemperatureAlarm}, if $\hatB$ leaves the ideal operational range after reaching once. 
\vspace{.2cm}
    \item [\textbf{\textit{FC5: }}] $A: \;TurnOffSignal(t) = \text{True} \;\; ,\;\; G:\;   \text{Burner} = \text{Off}$ 

    The burner must shut down immediately when it receives a \textit{TurnOffSignal} from the boiler.
\vspace{.2cm}
    \item [\textbf{\textit{FC6: }}] $A: \;\hatWM(t) > W_{{min}} \;\; ,\;\; G:\;   \hatWoM(t) > 0$
   
    The $\hatWoM$ must always remain above zero if the $\hatWM$ is above the minimum level, to keep the burner functional. 

\end{description}

\textbf{\textit{2- Boiler System FC:}}

The {\em boiler system} ensures stable operation by maintaining boiling conditions, monitoring water levels, and responding appropriately to critical conditions. 

\begin{description}
    \item [\textbf{\textit{FC7: }}] $
    A1:ReachedBoilingState(t)\; , A2: \hatWM(t) > W_{M_{min}} ,\; G: Bo_T(t) \approx T_{\text{Boil}} $
    
    If the boiler system has reached the boiling state and the $\hatWM$ is above the minimum level, the $\hatBo$ should remain around $T_{\text{Boil}}$.
\vspace{.2cm}
    \item [\textbf{\textit{FC8: }}] $
    A:\;\hatWM(t) < W_{M_{min}} \; ,\; G1:\;   W_A(t)\; , G2:\; TurnOffSignal(t)$
    
    If the $\hatWM$ falls below the minimum level, the system must generate $W_A$ and send a \textit{TurnOffSignal} to the burner to prevent unsafe operation.
\vspace{.2cm}
    \item [\textbf{\textit{FC9: }}] $
    A: \;\hatWM(t) > W_{M_{min}} \; , \;G1:\;   \neg W_A(t) \; , \;G2:\;\neg TurnOffSignal(t)$
    
    When the $\hatWM$ is above the minimum level, $W_A$ should not be triggered, and no \textit{TurnOffSignal} should be sent to the burner.
\vspace{.2cm}

    \item [\textbf{\textit{FC10: }}]$A1:\; \forall t, Burner = off \; , \;G1:\; \hatB(t) \geq T_{env}(t), \; , \;G2:\; \hatBo(t) \geq T_{env}(t)$

The $B_T$ and $Bo_T$ should always stay above or equal to the $T_{env}$, even if the burner is off. They may gradually decrease but should never drop below the ambient temperature, unless an external cooling source is applied. 
\end{description}
\subsection{Infrastructure Contracts (IC)}

\begin{description}
    \item [\textbf{\textit{IC1: }}] $
    A: \; \exists t, t_0 \text{ such that } (\text{CriticalAlarm}(t_0) = \text{True}) \wedge (\forall t' \in [t_0, t_0+5],$ $ \text{CriticalAlarm}(t') = \text{True}) \; , \;
    G: \; \text{ShutDownBurner}()
    $
    
    If a critical alarm remains active for more than $5 minutes$, the burner must shut down to prevent system damage and ensure safety.
\vspace{.2cm}

    %\item [\textbf{\textit{IC2: }}] $\text{BurnerOff}(t) \Rightarrow (\forall t' > t, \; Bo_T(t') - Bo_T(t'-1) \leq -\alpha \; \wedge \; W_T(t') - W_T(t'-1) \leq -\beta)$
    
    %When the burner is turned off, the boiler and water temperatures should gradually decrease by at least $\alpha$ and $\beta$ per time step, until they approach the environmental temperature $T_{env}$
\end{description}
%%%%%%%%%%%%%%%%%%%%%%%%%%%%%%%%%%%%%
\section{Contract Modeling in \uppaal}
\label{Modelling}
The UTA models presented in this section demonstrate how contract-based verification can check the correctness of a digital twin model, systematically. By mapping component and system contracts onto \uppaal timed automata, we ensure that key constraints are validated. \uppaal allows for efficient verification, as it explores all possible states and detects contract violations.
%%%%%%%
\subsection{Contract Model for MC1}
\label{Contract_Model_MC1}
\begin{figure}
    \centering
    \includegraphics[width=0.8\columnwidth]{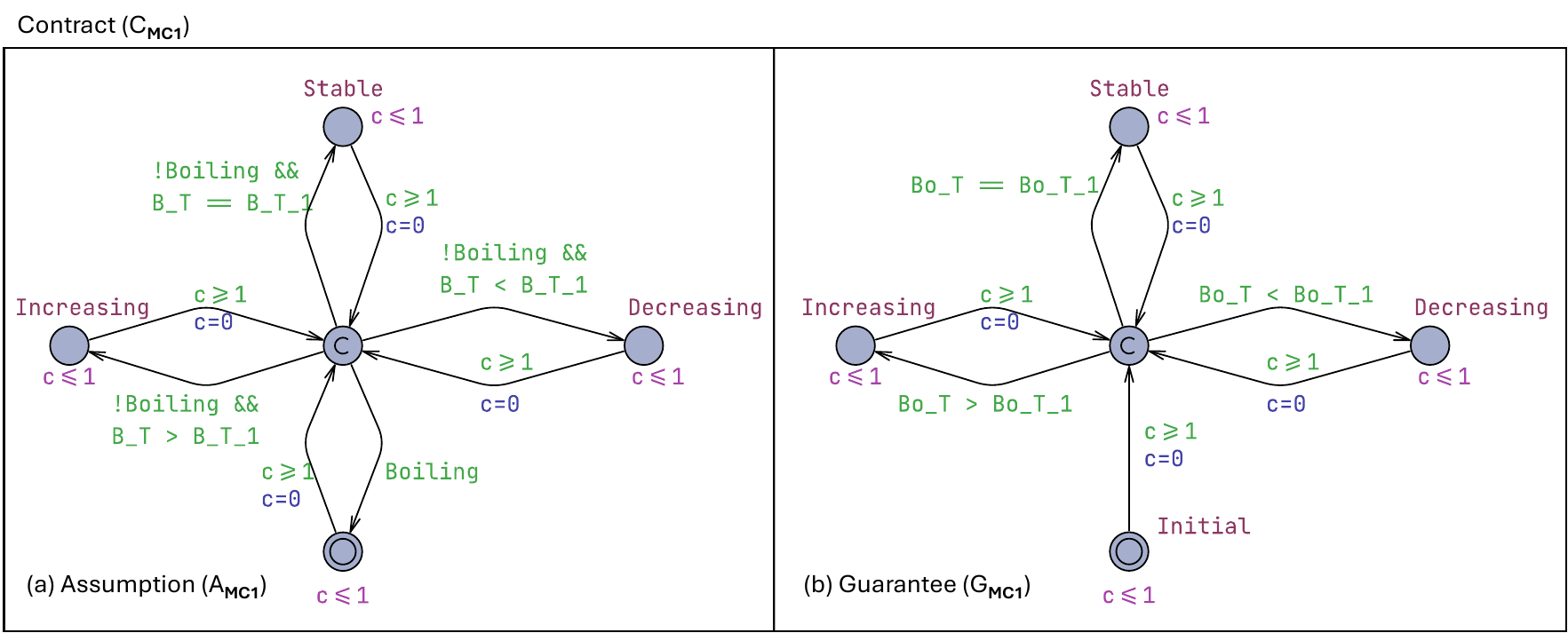}
    \caption{Monotonicity Contract UTA Model for MC1}
    \label{fig.Contract1}
\end{figure}

The Monotonicity Contract (MC1) ensures that the $Bo_T$ follows the changes in the $B_T$.  This contract guarantees a realistic and predictable behavior of the heating system.
The \uppaal model designed for MC1 is implemented as a timed automaton (Fig \ref{fig.Contract1}), effectively capturing the dynamic relationship between the temperatures. To model this contract in \uppaal, we used three templates: \textit{UpdateV} (Variable Update), $A_{MC1}$, and $G_{MC1}$. The \textit{UpdateV} template updates $B_T$ and $Bo_T$ values periodically. The $A_{MC1}$ template represents the expected behavior of the $B_T$, while the $G_{MC1}$ template ensures that the $Bo_T$v follows the $B_T$’s trend.

The \textit{UpdateV} template maintains the system’s temperature changes using global variables: $B_T$ (current burner temperature), $B_{T_{1}}$ (previous burner temperature), $Bo_T$ (current boiler temperature), and $Bo_{T_{1}}$ (previous boiler temperature). The function \textit{UpdateVar()} updates these values using a moving average. A clock variable \uppVar{(c)} ensures that updates occur at regular time steps.

$A_{MC1}$ template defines three possible states for the burner temperature: \uppLoc{Increasing} ($B_T > B_{T_{1}}$), \uppLoc{Decreasing} ($B_T < B_{T_{1}}$), and \uppLoc{Stable} ($B_T == B_{T_{1}}$). The system transitions between these states based on the changes in $B_T$. If the burner reaches the boiling point, the system enters a special state (Boiling).

The $G_{MC1}$ template enforces the expected reaction of the boiler temperature. It includes three states:  \uppLoc{Increasing} ($Bo_T > Bo_{T_{1}}$),  \uppLoc{Decreasing} ($Bo_T < Bo_{T_{1}}$), and  \uppLoc{Stable} ($Bo_T == Bo_{T_{1}}$). The transitions ensure that the boiler temperature behaves consistently with the burner temperature.

\subsubsection{Verification in \uppaal.}
\uppaal provides a query language to verify system properties \cite{behrmann2006tutorial}. The verification is done using queries, which check whether certain properties hold in all possible executions of the system.
To verify the correctness of the model, we define the following queries:

\textit{Safety Property Check}: This ensures that the system always satisfies the monotonicity requirement:
\begin{equation*}
    \uppAG{Assumption.Increasing imply Guarantee.Increasing}
\end{equation*}
If this query fails, it indicates that there are case(s) where the boiler temperature does not always follow the burner temperature correctly. Additionally, \uppaal provides diagnostic traces, which help locate errors by showing a counterexample when a property does not hold.
%/////////////////////////////////////////////////////
\subsection{Contract Model for FC9}
\label{Contract_Model_FC6}
The Functional Contract (FC9) ensures that when $W_M$ is above a defined minimum threshold, the alarm should not be triggered, and no turn-off signal should be sent to the burner. This contract prevents unnecessary interruptions in the heating system and ensures efficient operation.

\begin{figure}%[htbp]
    \centering
    \includegraphics[width=0.9\columnwidth]{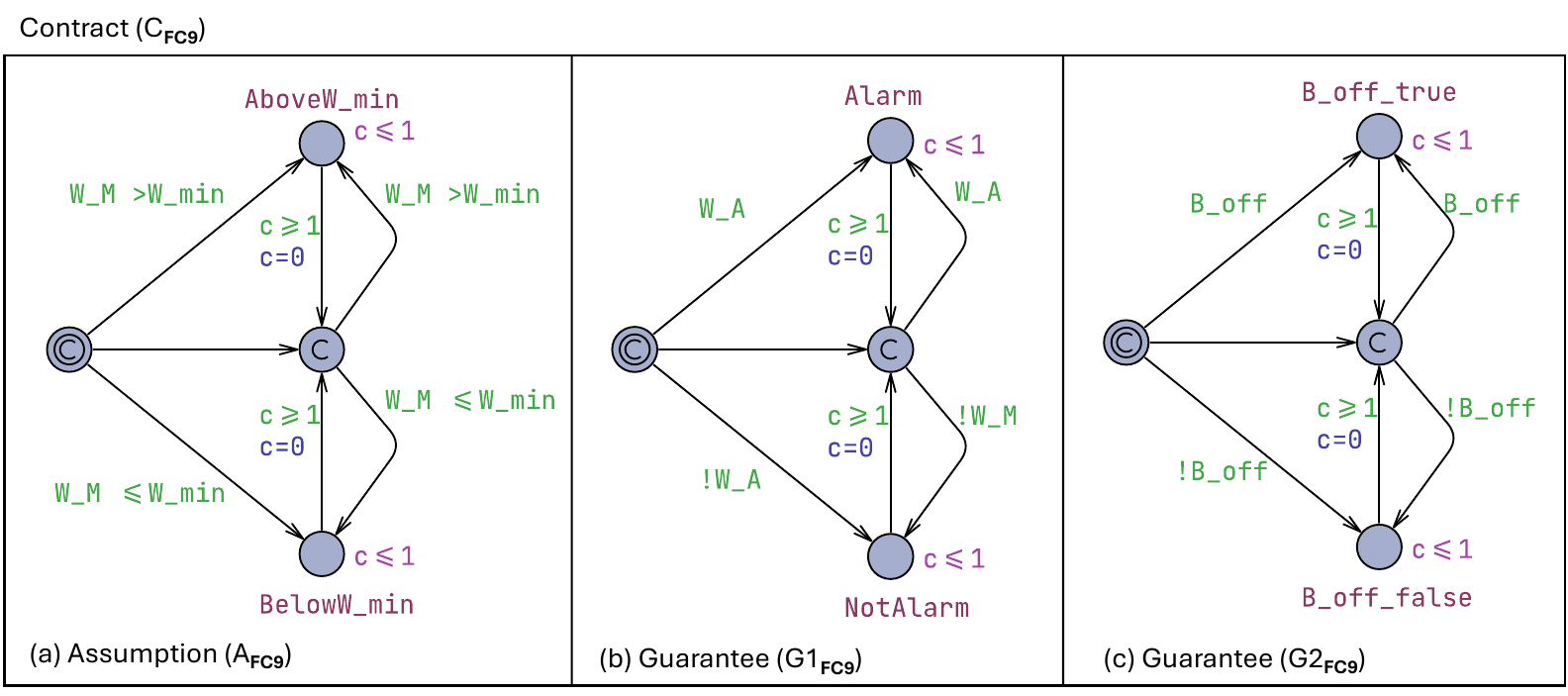}
    \caption{Functionality Contract UTA Model for FC9}
    \label{fig.Contract2}
\end{figure}

In Fig \ref{fig.Contract2}, the \uppaal model represents a contract (FC9) for a boiler system.
To model this contract in \uppaal, we define three key templates: $A_{FC9}$ (Water Level Monitoring), $G1_{FC9}$ (Alarm Condition), and $G2_{FC9}$ (TurnOffSignal). Additionally, a UpdateV (Variable Update) template periodically updates the monitored water level, similerly as described in MC1 Model. 

The $A_{FC9}$ template maintains the state of the water level, defining two primary locations: \uppLoc{AboveW_min}, and \uppLoc{BelowW_min}, where . A clock variable (c) ensures regular updates, and the system transitions between these states based on the water level condition.

The $G1_{FC9}$ template models the alarm behavior by defining two states: \uppLoc{NotAlarm}, and \uppLoc{Alarm}. The transition to the Alarm state occurs only if the water level falls below the threshold. If the water level is above the threshold, the system remains in the NotAlarm state, ensuring that the contract requirement is met.

The $G2_{FC9}$ template models the burner control system by ensuring that the burner remains on as long as the water level is above the minimum threshold. The states in this template include \uppLoc{B_off_false}, and \uppLoc{B_off_true}. The transition to \uppLoc{B_off_true} is allowed only if, ensuring that the burner is not turned off unnecessarily.

\subsubsection{Verification in \uppaal.}
To verify the correctness of the model, we define the following queries. The \textit{invariance property check} ensures that, when the water level is above the threshold, neither the water alarm is triggered, nor the burner is turned off. This is expressed as:
\begin{equation*}
    \uppAG{A_FC9.AboveW_min  imply  (not (G1_FC9.Alarm || G2_FC9.B_off_true))}
\end{equation*}

If this property fails, it indicates that $W_A$ or {\em turn-off} signal is incorrectly triggered despite the water level being sufficient. The reachability check verifies whether a violation of the contract is possible:
\begin{equation*}
\uppEF{ A_FC9.AboveW_min  imply  (G1_FC9.Alarm || G2_FC9.B_off_true)}
\end{equation*}
%%%%%%%%%%%%%%%%%%%%%%%%%%%%%%%%%%%%%
\section{Verification Results}
\label{verification}

Our contract-based verification approach using UPPAAL revealed some scenarios in which the predictions of the NNDT model deviated from the specified contracts. These findings highlight areas for improvement in the model and potential safety concerns regarding the boiler system's operation.

\subsection{MC1: Relation between the burner and the boiler temperatures}
We simulate the contract automaton model described in section \ref{Contract_Model_MC1} to verify the relation between the burner and the boiler temperatures. 
The results illustrate a violation of the expected relationship between burner and boiler temperatures. According to MC1, the boiler temperature should correlate with the burner temperature, allowing for an acceptable delay $\delta$, particularly when the burner temperature is below the boiling threshold.
\begin{figure}
    \centering
    \includegraphics[width=0.6\columnwidth]{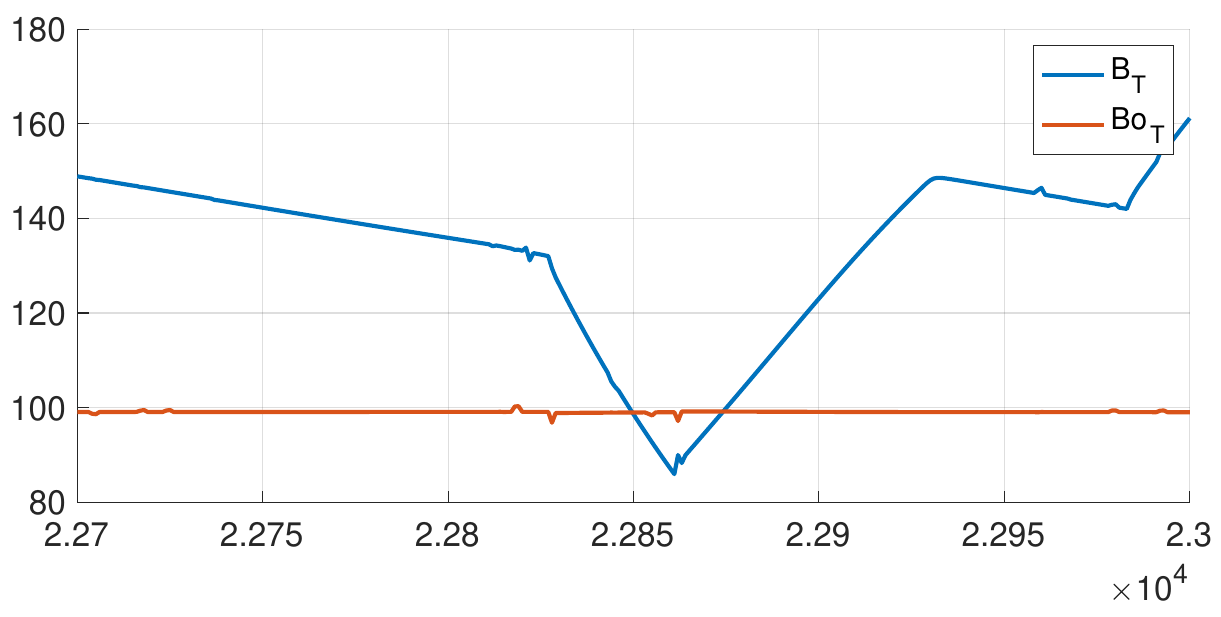}
    \caption{Contract violation in MC1}
    \label{fig.MC_violation}
\end{figure}

In Figure \ref{fig.MC_violation}, the graph shows that the burner temperature gradually decreases with time, even dropping below the boiling temperature at certain points. Despite this, the boiler temperature remains unchanged and does not follow the expected correlation. This discrepancy suggests that the heat transfer dynamics between the burner and the boiler are not accurately trained in the digital twin model.

The impact of this contract breach is significant. It may result in erroneous system behavior, such as the boiler maintaining a high temperature despite a decrease in burner temperature, which could lead to incorrect operational conditions. This conclusion emphasizes the need for model refinement to ensure that the digital twin accurately represents temperature dependencies and state transitions, enhancing its fidelity and reliability as a predictive tool.
\subsection{FC3: Erroneous wood request}
\begin{figure}
    \centering
    \includegraphics[width=0.7\columnwidth]{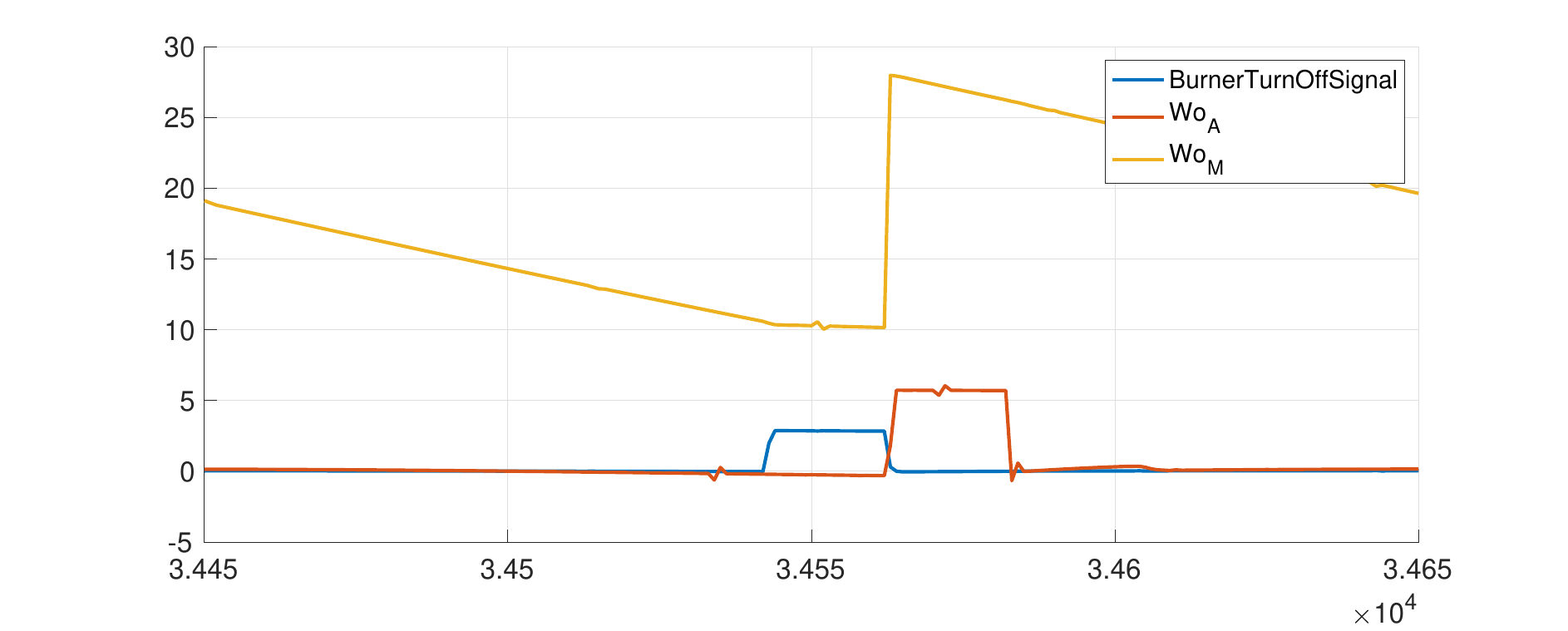}
    \caption{Contract violation in FC3}
    \label{fig.FC3_violation}
\end{figure}

The results indicate a contract violation in the wood request mechanism, leading to an unnecessary wood supply. According to FC3, a wood request should only be triggered when the wood mass falls below the minimum threshold. However, as shown in Figure \ref{fig.FC3_violation}, noise in the wood request signal causes the system to falsely detect a low wood level, even when there is enough wood available.

As a result, the model requests unnecessarily additional wood, leading to an excessive amount of wood in the burner. This misbehavior can affect system efficiency and fuel management, potentially causing overheating or inefficient combustion.

Improvements in the wood request signal processing are required to avoid such errors. Possible solutions include noise-filtering techniques or threshold adjustments to avoid false triggers and ensure accurate wood requests.
%%%%%

These verification results highlight the importance of refining the DT model, particularly in areas of wood mass management, temperature control, and system state consistency. They also demonstrate the effectiveness of contract-based verification in identifying potential safety and operational issues that might not be apparent through traditional testing methods.
\subsection{FC9: Alarm Water Finished False Trigger}
\begin{figure}
    \centering
    \includegraphics[width=0.7\columnwidth]{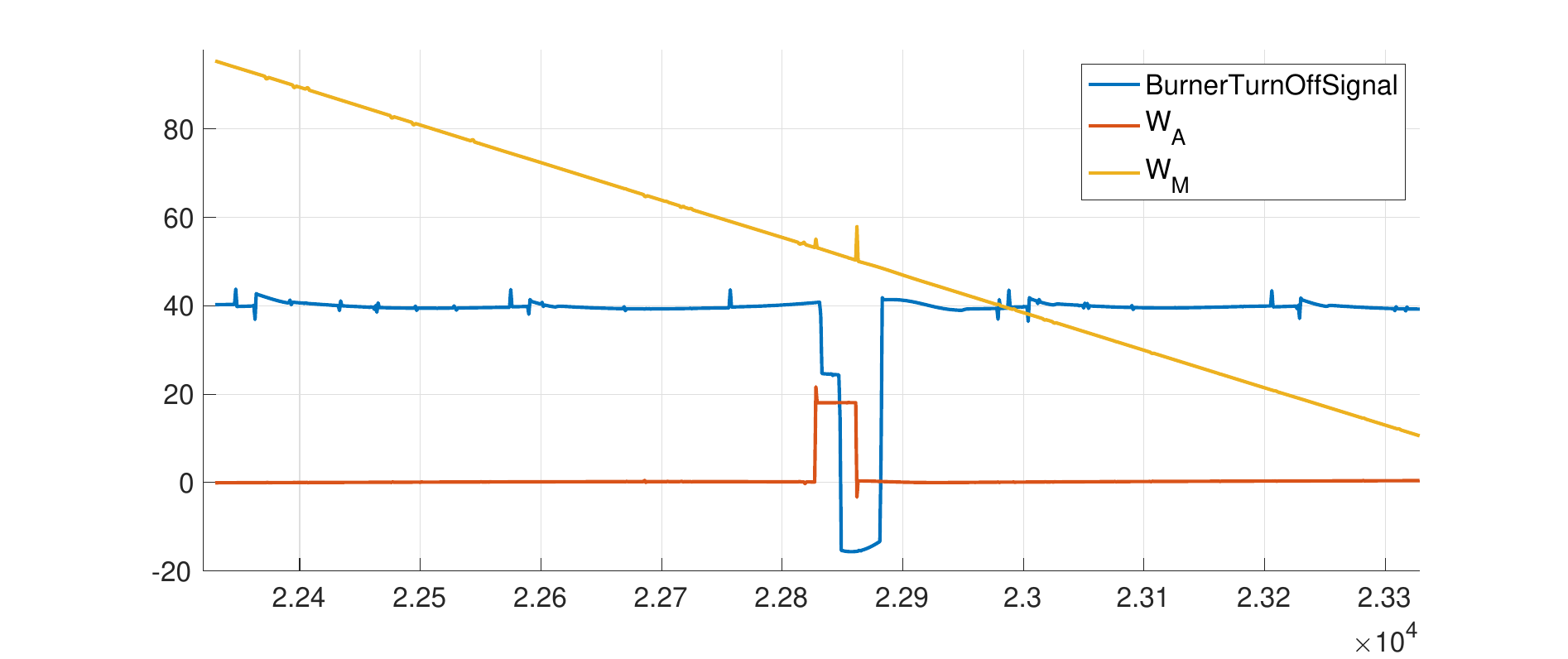}
    \caption{Contract violation in FC9}
    \label{fig.FC9_violation}
\end{figure}
The findings highlight a contract violation in the Water\_Finished\_Alarm, which led to an incorrect shutdown of the burner. According to FC9, the alarm should activate only when the water mass falls below the minimum threshold. However, as illustrated in figure \ref{fig.FC9_violation}, interference in the alarm signal leads to a false burner shutdown, even when the water mass remains above the minimum level. 

Improvements in alarm signal processing are necessary to address this issue. Possible solutions include implementing signal filtering techniques, adjusting threshold conditions, or introducing hysteresis mechanisms to prevent minor fluctuations from causing false shutdowns.
%%%%%%%%%%%%%%%%%%%%%%%%%%%%%%%%%%%%%
\section{Related Work}
\label{RW}

The verification of DT is a growing area of research due to its increasing use in critical applications. 

Cimatti et al. provide the OCRA tool~\cite{Cimatti13} that enables a {\em design-by-contract} methodology for reactive systems. In OCRA, the system's DT is its contract, which consists of an assumption on the system inputs, and a guarantee of the system outputs under the given assumptions, both described in Linear Temporal Logic. In comparison, our approach distinguishes between the DT, which is a neural network, and its contracts, and specifies the latter as timed automata that are fed with the DT's inputs and predicted outputs, to verify the black-box behavior of the DT by model checking. 

Perhaps the closest research to ours is the approach used in CoCoSim~\cite{bourbouh2020cocosim}, which is a toolbox that can be called directly from the Matlab Simulink environment, and can be used for code generation (e.g. Lustre) or property verification via contracts. The contracts are specified directly in Simulink, as model components, and the verification is carried out on the generated code, by using modular SMT-based verification engines (such as Zustre) that also generate assume-guarantee style formal contracts. However, the difference from our approach comes from the fact that the verification is not a black-box one, it instead uses the functional details of Simulink's model blocks that are transformed into code. The fact that our approach uses contracts modeled as simple timed automata that encode the properties to be model checked makes the verification very fast and scalable. 

Notable work for {\em verifying neural networks}, which in our case describes the DT, has been carried out for a while now. Katz et al. introduce the ``Reluplex'' method~\cite{Katz17}, designed to verify deep neural networks by solving satisfiability problems, particularly in networks utilizing ReLU activation functions. In comparison, our method is agnostic of the NNDT's activation function, being based only on inputs, predicted outputs and their check of whether they satisfy the timed automata contracts. 

Another strand of research focuses on {\em reachability analysis}, where methods such as {\em symbolic interval propagation}~\cite{Kern22} and {\em abstract interpretation}~\cite{Singh19,Tran20,Lemesle24,Gehr18,Mazzucato21} are used to estimate possible outputs of neural networks within a bounded input space. This is crucial for safety-critical systems to ensure that neural networks do not produce unexpected or unsafe outputs. For instance, the Neural Network Verification tool (NNV)~\cite{Tran20} supports over-approximate analysis by combining the star set analysis used for feed-forward neural network controllers with zonotope-based analysis for nonlinear plant dynamics. In general, neural network verification approaches based on abstract interpretation are supported by tools that represent the inputs as a set, called abstract domain, which is then passed along the neural network, yielding a set that over-approximates the output that is used to evaluate the property under verification. The verification accuracy depends heavily on the abstract domain chosen~\cite{Lemesle24}. If the abstraction is too rough, the tool might not be able to decide if the property is verified or falsified, therefore other methods need to be employed. By using contracts that encode the actual NNDT properties that need to be verified, we avoid such problems, providing a fast and scalable solution that can potentially be employed at run-time too, due to its insignificant overhead.  

Additionally, researchers explore {\em probabilistic verification}~\cite{Boetius24} to handle the inherent uncertainty in data-driven models like neural networks. Techniques based on Bayesian methods and Monte Carlo simulations have been proposed to evaluate the performance and robustness of neural networks DT in dynamic environments. Kapteyn et al.~\cite{Kapteyn21} explore the mathematical foundations of digital twins, focusing on probabilistic graphical models, including dynamic Bayesian networks, to model complex systems and inform decision-making processes.
%%%%%%%%%%%%%%%%%%%%%%%%%%%%%%%%%%%%%
\section{Conclusions and Future Work}
\label{conclusion}
This paper presents a novel methodology for verifying black-box neural network-based digital twin models using contract-based model checking. Our approach, which uses \uppaal timed automata to define system contracts, allows for black-box verification without requiring knowledge of the digital twin's internal details. This makes it particularly valuable for complex industrial applications.
We validated our methodology using an existing neural network-based digital twin model of a boiler system developed in Simulink. By modeling contracts as \uppaal timed automata, we demonstrated how our approach can be applied to real-world industrial processes, ensuring digital twin predictions align with physical constraints and expected behaviors.

Our contract models show fast simulation times, making them suitable for runtime verification. This efficiency enables real-time monitoring of the digital twin's behavior during system operation without significantly impacting overall performance.

Future work will focus on analyzing the digital twin's performance under various input conditions and exploring ways to improve predictions for more complex systems. We aim to enhance the reliability and utility of digital twins in industrial applications, leading to improved efficiency and safety in operations.
%%%%%%%%%%%%%%%%%%%%%%%%%%%%%%%%%%%%%
\begin{credits}
\vspace{-.2cm}
\subsubsection{\ackname}
The research was supported by vinnova’s advanced digitalization program in the project D-RODS (ID: 2023-00244).
\end{credits}
%%%%%%%%%%%%%%%%%%%%%%%%%%%%%%%%%%%%%
\bibliographystyle{splncs04}
\bibliography{main}

\end{document}